\documentclass[12pt,a4paper]{article}
\usepackage{amsmath}
\usepackage{amssymb}
\usepackage{graphicx}
\begin{document}
\begin{titlepage} 
\begin{flushright} IFUP--TH/2025\\
\end{flushright} ~\vskip .8truecm                 

~ 
\vskip 3.0truecm

\begin{center} 
\Large\bf Convergence of classical conformal blocks
\end{center}
\vskip 1.2truecm
\begin{center}
{Pietro Menotti} \\   
{\small\it Dipartimento di Fisica, Universit{\`a} di Pisa}\\ 
{\small\it 
Largo B. Pontecorvo 3, I-56127, Pisa, Italy}\\
{\small\it e-mail: pietro.menotti@unipi.it}\\ 
\end{center} 
\vskip 0.8truecm
\centerline{December 2025}
\vskip 1.2truecm
                                                              
\begin{abstract}
  We give a recursive method to compute the classical conformal blocks
  in Liouville field theory. The values of the expansion coefficients
  are given by an algebraic scheme which works to all orders. The
  algebraic expression of the intervening matrices are explicitly
  given. With regard to the problem of the convergence of the series
  we rigorously prove that it has a finite (non zero) convergence
  radius.
  We then comment on the relation of the conformal block problem with
  the Riemann-Hilbert problem.
\end{abstract}

\end{titlepage}

\section{Introduction}\label{introduction}
A lot of attention has been devoted to the conformal blocks both at
the quantum level and in the classical limit
\cite{ZZ}-\cite{LN}
For the four point function the conformal blocks are given as a formal
series in the modulus $x$ of the problem \cite{AlZ1,AlZ2,DMS}.

In the classical limit, i.e. $b\rightarrow 0$ the quantum conformal
blocks exponentiate \cite{ZZ} with the classical block at the exponent
multiplied by $b^{-2}$. Such fact is not at all trivial as in the
process heavy cancellations occur. The derivative of the classical
conformal block w.r.t. the modulus is simply connected with the
accessory parameter of an Heun differential equation for which a given
trace, i.e. the class, of a particular monodromy is realized
\cite{LLNZ}.

In \cite{sphere4,sphere5} a straightforward algebraic method was
developed to compute explicitly the coefficients of the series
expressing the classical conformal blocks. The extension of the
technique to the torus topology was given in \cite{menottitorus}.
Such procedure works at all orders and has been successfully compared
with the classical limit of the quantum conformal blocks available in
the literature \cite{MMM,LLNZ,ferraripiatek}.

An important question is whether such a series expressing the classical
conformal blocks is just a formal, possibly asymptotic series, or it
is a convergent series in which case one is interested in the
radius of convergence of the series.

In the algebraic approach given in section \ref{algebraic}
a key role is played by a nested
family of non singular upper triangular matrices and their
inverses. We give the explicit form of them and of their inverses. We
provide also an upper bound to the norm of such inverse matrices.  In
order to establish the convergence of the series we turn to the non
perturbative expression of the monodromy matrices and thus of their
traces. This is done with the Green function method. We find that the
series for the trace converges in a disk of radius $1$ around
zero. This however does not give directly the value of the related
accessory parameter which is given by the solution of an implicit
equation. Using tools of analytic varieties \cite{whitney} we prove
that the series in the modulus converges in a finite disk around the
origin and we give a procedure to establish a lower bound to the
convergence radius which obviously depends on the parameters of the
theory.

One should not confuse the accessory parameter $C$ appearing in the
present problem, which as we shall see, at least in a neighborhood of
$x=0$ depends analytically on $x$, with the accessory parameter
$C(x,\bar x)$ which appears in the auxiliary equation associated to
the solution of the Liouville equation, related to the uniformization
problem, which is not an analytic function of $x$ but a real analytic
function of such a variable i.e. it is the value of an analytic
function of two variables $C(u,v)$ when $u=x$ and $v=\bar x$. The real
analytic dependence of $C$ was proven in
\cite{kra,menottielliptic,menottiparabolic}. The two however are
related as was shown in \cite{ZZ}.

As stressed in \cite{LLNZ} such a problem is of different nature of
the one related to the uniformization
\cite{picard1}-\cite{menottisolution}.
In fact one expects, and will be proven in
sections \ref{green}, the dependence of $C$ on the modulus $x$ to be
analytic, while in the uniformization case as proven in
\cite{menottielliptic,menottiparabolic} the dependence is real
analytic.

The structure of the paper is the following: In section
\ref{classicallimit}
we recall
the role of conformal blocks in building up the four point function
and its relation to the accessory parameter appearing in the auxiliary
equation related to the solution of the Liouville equation.

In section \ref{algebraic} we give the algebraic treatment. With
respect to paper \cite{sphere5} the notation has been improved and the
fundamental matrix $A^{(N)}$ is given explicitly to all orders.  In
section \ref{green} we give the complete form of the trace of the
monodromy appearing in the problem. The series in $x$ converges
rigorously for all $|x|<1$ and for any $C$ and thus the result is non
perturbative. We prove that the ensuing function $C(x)$ which solves
the equation for ${\rm tr} M_{0x}$ is analytic in $x$ in a disk of
finite radius giving also a procedure to determine a lower bound on
the convergence radius.  We then compute the first and second order
which coincide with the results of the algebraic approach of the
previous section and prove that such agreement extends to all orders
in $x$.  In section \ref{riemannhilbert} we discuss the relations of
the treated problem with the classical Riemann-Hilbert problem
\cite{bolibrukh}.  In section \ref{conclusions} we give the conclusions
and point out some open problems.

\section{The classical limit}\label{classicallimit}

In this section we review the relations between the accessory
parameter which appears in the auxiliary equation related to the
regular solution of the Liouville equation
\begin{equation}\label{liouvilleeq}
-\frac{1}{4}\Delta \phi(z) + e^{\phi(z)}={\rm sources} 
\end{equation}
and the accessory parameter which appear in the monodromy problem
of LLNZ \cite{LLNZ}.
The quantum four point function can be written as \cite{ZZ,HJP}
\begin{equation}\label{ZZdecomp}
  G = \frac{1}{2}\int_{-\infty}^\infty dP~
  C(\alpha_1,\alpha_2,\frac{Q}{2}+iP)~
  C(\alpha_3,\alpha_4,\frac{Q}{2}-iP)~
  |{\cal F}(\Delta_i,\Delta,x)|^2
\end{equation}
\begin{equation}
  \alpha = \frac{Q}{2}+iP,~~~~\Delta = \frac{Q^2}{4}+P^2,~~~~Q=
  b+\frac{1}{b}
\end{equation}
where $C(\alpha_1,\alpha_2,\alpha)$ is the quantum three point function
and ${\cal F}$ is the conformal block.  In the classical limit
$b\rightarrow 0$, $\alpha_i=\eta_i/b$ we know that the $C$ goes over
to
\begin{equation}
  C(\alpha_1,\alpha_2,\alpha_3)~\sim~{\rm exp}\bigg(-\frac{1}{b^2}
  S^{(cl)}(\eta_1, \eta_2,\eta_3)\bigg) 
\end{equation}
\begin{equation}
  \alpha_j = \frac{\eta_j}{b}, ~~~~~~~~\Delta_j= \alpha_j(\frac{1}{b}+b
  -\alpha_j)
  \rightarrow \frac{\delta_j}{b^2}
\end{equation}
where $S^{(cl)}$ is the well known three point classical Liouville
action.  The classical conformal field $\phi(z)$ which solves
(\ref{liouvilleeq}) is given by \cite{CMS2}
\begin{equation}
e^{-\frac{\phi}{2}} = \bar y_2 y_2 - \bar y_1 y_1
\end{equation}
where $y_1,y_2$ are two independent solutions of the auxiliary ODE
\begin{equation}\label{accessoryeq}
y''+Q y=0,~~~~ Q 
=\frac{\delta_0}{z^2}+\frac{\delta}{(z-x)^2}+\frac{\delta_1}{(z-1)^2}
+\frac{\delta_\infty-\delta_0-\delta-\delta_1}{z(z-1)}
-\frac{x (1-x)C_L}{z(z-x)(z-1)}~.
\end{equation}  
The $C_L$ which appear in eq.(\ref{accessoryeq}) is the accessory
parameter which has to be chosen so that the $\phi$ is single valued
in the $z$ plane.  $C_L$ is related to the classical action by the
Polyakov relation \cite{CMS1,CMS2,TZ}
\begin{equation}\label{polyrel}
C_L(x,\bar x) = -\frac{\partial S(x,\bar x)}{\partial x}
\end{equation}
where $S(x,\bar x)$ is the on shell Liouville action.  $C_L$ is not an
analytic function of the modulus $x$ but a real analytic function of $x$
i.e. the value which an analytic function of two variables $C_L(u,w)$
assumes for $u=x$ and $w=\bar x$ and it solves the uniformization
problem \cite{kra,menottielliptic,menottiparabolic}.

The conformal block ${\cal F}$ of eq.(\ref{ZZdecomp}) in the limit
$b\rightarrow 0$ exponentiates to \cite{ZZ}
\begin{equation}
{\cal F}~\sim~ {\rm exp}\bigg(\frac{1}{b^2}f(\eta_i,p,x)\bigg)~.
\end{equation}
Such an exponentiation in highly non trivial as in the process heavy
cancellations occur. 

The occurrence of the factor $\frac{1}{b^2}$ means that the classical
four point function can be written as the exponential of $-1/b^2$ times
\begin{equation}
  S^{(cl)}(\eta_1,\eta_2,\eta_3,\eta_4,x,\bar x)
  =S^{(cl)}(\eta_1,\eta_2,\frac{1}{2}+ip_s)+
  S^{(cl)}(\eta_3,\eta_4,\frac{1}{2}-ip_s)
  -f(\eta_i|p_s,x)-f(\eta_i|p_s,\bar x)
\end{equation}
where $p_s$ is the saddle point of the integral (\ref{ZZdecomp}) over
$p=\frac{P}{b}$ i.e. the value of $p$ where, after defining
\begin{equation}\label{saddlepoint}
 {\cal S}_{\eta_1\eta_2\eta_3\eta_4}(p|x,\bar x)=
  S^{(cl)}(\eta_1,\eta_2,\frac{1}{2}+ip)+
  S^{(cl)}(\eta_3,\eta_4,\frac{1}{2}-ip)
-f(\eta_i|p,x)-f(\eta_i|p,\bar x)
\end{equation}
we have
\begin{equation}\label{saddlepoint2}
\frac{\partial}{\partial p} S_{\eta_1,\eta_2,\eta_3,\eta_4}(p|x,\bar x)=0~.
\end{equation}

Then applying eq.(\ref{saddlepoint}) and using eq.(\ref{polyrel}) we
have
\begin{equation}\label{CLxbx}
C_L(x,\bar x)= \frac{\partial}{\partial x}f(\eta_i,p_s,x)~.
\end{equation}
Notice that one expects $f(\eta_i,p,x)$ for fixed $p$ to be, apart a
logarithmic term \cite{ZZ}, a power series in $x$ and thus if such a
series is convergent, as we shall prove, an analytic function of $x$
for $x$ near $0$.  The non analytic but real analytic nature of the
accessory parameter $C_L$ is due to the presence in eq.(\ref{CLxbx})
of $p_s$ which due to (\ref{saddlepoint2}) depends both on $x$ and
$\bar x$.  In fact accessory parameters related to the uniformization
problem were rigorously proven to be real analytic functions of the
position of the sources \cite{kra,menottielliptic,menottiparabolic}
except when two sources coalesce; in particular $C_L(x,\bar x)$ is
singular at $x=0$ \cite{HS2}.

\section{The algebraic approach}\label{algebraic}

In presence of four singularities the differential equation takes the
form
\begin{equation}\label{fundeq}
y''(z)+ Q(z) y(z)=0
  \end{equation}
with
\begin{equation}\label{fundQ}
  Q(z) =\frac{\delta_0}{z^2}+\frac{\delta}{(z-x)^2}+\frac{\delta_1}{(z-1)^2}
  +\frac{\delta_\infty-\delta_0-\delta-\delta_1}{z(z-1)}
  -\frac{C(x)}{z(z-x)(z-1)}~.
\end{equation}
where we use the notation $C= x(1-x)C_L$.

As already discussed in the introduction, we are faced with the
following problem \cite{LLNZ}: Given the class of the monodromy of the
circuit embracing $0$ and $x$, i.e. given
\begin{equation}
{\rm tr}M_{0x} = -2\cos\pi\lambda_\nu,~~~~\delta_\nu=\frac{1-\lambda^2_\nu}{4}
\end{equation}
determine the value of $C(x)$ for which such a value is realized for
the pair of solution of (\ref{fundeq}).

We are in particular interested in the nature of the dependence of
$C(x)$ on the ``modulus'' $x$.

For $x=0$ we have $C(0)=\delta_\nu-\delta_0-\delta$
and $Q(z)$ goes over to
\begin{equation}
  Q_0(z)=\frac{\delta_\nu}{z^2}+\frac{\delta_1}{(z-1)^2}+
  \frac{\delta_\infty-\delta_1-\delta_\nu}{z(z-1)}~.
\end{equation}
In order to compute $C(x)$ as a power expansion in $x$ which is the usual
presentation of the conformal blocks, we expand (\ref{fundeq}) in $x$
reaching
\begin{equation}
  Q = Q_0+ x Q_1+x^2 Q_2+\cdots
\end{equation}
with
\begin{equation}\label{Qexpansion}
Q_n=\frac{Q^{(n)}}{n!}= 
\frac{1}{z(z-1)}\bigg[\frac{-(n+1)\delta-C(0)}{z^{n+1}}+
\frac{(n+1)\delta-C'(0)}{z^n}-
\sum_{k=0}^{n-2}\frac{C^{(n-k)}(0)}{(n-k)!}\frac{1}{z^{1+k}}\bigg]~.
\end{equation}
\bigskip

We know from the theory of ODE \cite{yoshida} that if we perform the
change of variable $z=z(v)$ the equation
\begin{equation}\label{Q0equation}
y''(z)+ Q_0(z) y(z)
\end{equation}
goes over to
\begin{equation}\label{vequation}
y^{A''}(v)+ Q_v(v) y^A(v)=0
\end{equation}
with
\begin{equation}\label{Qv}
  Q_v(v)= Q_0(z(v))\bigg(\frac{dz}{dv}\bigg)^2 -\{z,v\}~~~~{\rm and}~~~~
    y^A(v) = y(z(v))\bigg(\frac{dz}{dv}\bigg)^{-\frac{1}{2}}~.
\end{equation}
The Schwartz derivative $\{z,v\}$ is given by
\begin{equation}
\{z,v\} =
(\frac{dz}{dv})^{\frac{1}{2}}\frac{d^2}{dv^2}(\frac{dz}{dv})^{-\frac{1}{2}}~.
\end{equation}
A useful relation for computing the Schwarz derivative is its behavior
under a change of variables (see e.g. \cite{yoshida})
\begin{equation}
\{f\circ g,x\} = \{f,g\}(\frac{dg}{dx})^2+\{g,x\}~.
\end{equation}
The main idea of \cite{sphere4,sphere5} is to compute order by order
in $x$ the monodromy along the circuit I of figure.1 by deforming it
to the equivalent contour II.
\bigskip\bigskip\bigskip
\begin{figure}[htb]
\begin{center}
\includegraphics{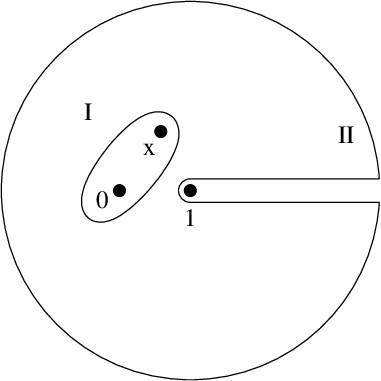}
\end{center}
\caption{The monodromy contour}
\end{figure}

For $x=0$ the solutions of (\ref{Q0equation}) are known in terms of
hypergeometric functions \cite{sphere4,sphere5} and their asymptotic
behaviors for $z=+\infty$ above the cut are given by
\begin{eqnarray}\label{Y+}
Y_0^+(z)&=&
\begin{pmatrix}
-ie^\frac{i\pi\lambda_1}{2} &0\\
0 &-ie^{-\frac{i\pi\lambda_1}{2}}
\end{pmatrix}
\begin{pmatrix}
t_1(z)\\
t_2(z)
\end{pmatrix}
\equiv \Lambda_1
\begin{pmatrix}
t_1(z)\\
t_2(z)
\end{pmatrix}\nonumber\\
&\approx& \Lambda_1 B
\begin{pmatrix}
z^\frac{1-\lambda_\infty}{2}\\
z^\frac{1+\lambda_\infty}{2}
\end{pmatrix}
\equiv B^+\begin{pmatrix}
z^\frac{1-\lambda_\infty}{2}\\
z^\frac{1+\lambda_\infty}{2}
\end{pmatrix}~.
\end{eqnarray}
Similarly below the cut we have
\begin{equation}\label{Y-}
Y_0^-(z)= \Lambda_1^{-1}
\begin{pmatrix}
t_1(z)\\
t_2(z)
\end{pmatrix}
\approx 
\Lambda^{-1}_1B
\begin{pmatrix}
z^\frac{1-\lambda_\infty}{2}\\
z^\frac{1+\lambda_\infty}{2}
\end{pmatrix}\equiv
B^-
\begin{pmatrix}
z^\frac{1-\lambda_\infty}{2}\\
z^\frac{1+\lambda_\infty}{2}
\end{pmatrix}~.
\end{equation}
 From the
asymptotic behaviors (\ref{Y+},\ref{Y-}) one has that the monodromy
matrix associated to the contour II of fig.1 is
\begin{equation}
M^0= B^+\Lambda_\infty (B^-)^{-1} 
\end{equation}
where 
\begin{equation}
\Lambda_\infty=
\begin{pmatrix}
e^{i\pi(1-\lambda_\infty)}&0\\
0&e^{i\pi(1+\lambda_\infty)}
\end{pmatrix},
\end{equation}
\begin{equation}\label{Bplusminus}
B^+ = \Lambda_1 B,~~~~~~~~B^- = \Lambda_1^{-1}B ~,   
\end{equation}
with
\begin{equation}
\Lambda_1=
\begin{pmatrix}
-i e^{\frac{i\pi\lambda_1}{2}}&0\\
0&-i e^{-\frac{i\pi\lambda_1}{2})}
\end{pmatrix}~.
\end{equation}
The matrix $B$ is know, see eq.(\ref{Bmatrix}), and from its explicit
value one can verify that the trace of the monodromy is $-2 \cos
\pi\lambda_\nu$ as it must be. However the explicit value of $B$ will
not be necessary for the following algebraic developments. 

We consider now on eq.(\ref{Q0equation}) the transformation of variable
\begin{equation}\label{ztovchange}
  z(v) = \frac{v-{\cal B}_0-{\cal B}_1/v-{\cal B}_2/v^2+\cdots}
        {1-{\cal B}_0-{\cal B}_1-{\cal B}_2+\cdots}
\end{equation}
where
\begin{eqnarray}\label{Bexp}
  &&  {\cal B}_0= x b_{0,1}+x^2 b_{0,2}+x^3 b_{0,3}+\cdots\nonumber\\
  &&  {\cal B}_1= x^2 b_{1,1}+x^3 b_{1,2}+x^4 b_{1,3}+\cdots\nonumber\\
  &&  {\cal B}_2= x^3 b_{2,1}+x^4 b_{2,2}+x^5 b_{2,3}+\cdots\\
  &&  {\cal B}_3= x^4 b_{3,1}+x^5 b_{3,2}+x^6 b_{3,3}+\cdots\nonumber\\
  &&   .... \nonumber
\end{eqnarray}
Given a fixed order $N$ in $x$ we remark that the point $z=1$ and
$z=\infty$ are fixed under the transformation (\ref{ztovchange}) and
it is simple to prove \cite{sphere5} that given any $r$ with $0<r<1$
and $|v|>r$ for $|x|$ small enough the relation $z(v,x)$ is one to
one. The monodromy contour II of fig.1 lies in this region.

Thus in computing the monodromy matrix the only thing that changes
with respect to the previous calculation are the matrices $B^+$ and
$B^-$ which change by the right multiplication with a diagonal matrix
$D$.  In fact for large $v$ we have
\begin{equation}
  z(v) \approx \frac{v}{1-{\cal B}_0-{\cal B}_1\cdots}
  \equiv \frac{v}{1-{\cal B}_T}
\end{equation}
and  
\begin{equation}
  y^A_j(v) = y(z(v)) (z'(v))^{-\frac{1}{2}} \approx y_j(\frac{v}{1-{\cal B}_T})
  (1-{\cal B}_T)^\frac{1}{2}~.
\end{equation}

Then we have
\begin{equation}
  Y^{A+}(v) \approx \Lambda_1 B D
  \begin{pmatrix}
    v^{\frac{1-\lambda_\infty}{2}} \\
    v^{\frac{1+\lambda_\infty}{2}}
  \end{pmatrix}
\end{equation}
where
\begin{equation}
  D=
  \begin{pmatrix}
    (1-{\cal B}_T)^\frac{\lambda_\infty}{2}&0\\
    0&(1-{\cal B}_T)^{-\frac{\lambda_\infty}{2}}
  \end{pmatrix}
\end{equation}
and similarly
\begin{equation}
  Y^{A-}(v) \approx {\Lambda_1}^{-1}B  D
  \begin{pmatrix}
    v^{\frac{1-\lambda_\infty}{2}} \\
    v^{\frac{1+\lambda_\infty}{2}}
  \end{pmatrix}
\end{equation}
and thus the new monodromy becomes
\begin{equation}
  M= \Lambda_1 B D \Lambda_\infty D^{-1} B^{-1} \Lambda_1=
  M^0
\end{equation}
due to the diagonal nature of $\Lambda_\infty$ and $D$
i.e. the monodromy is unchanged.

Thus in the algebraic approach one expresses the solutions of the
problem by means of a particular change of variable in the unperturbed
solution $Y_0$ which assures an unchanged trace of the monodromy. Now
we impose that such a function solves the equation (\ref{fundeq})
i.e. we equate term by term the expansion in powers of $x$ of $Q(v)$
of (\ref{fundeq}) with the $Q_v(v)$ given by (\ref{Qv}) i.e.
\begin{equation}\label{Qequation}
Q(v,x) = Q_0(z(v,x)) (z'(v,x))^2 -\{z(v,x),v\}~.
\end{equation}

To 
order $x^0$ we obtain
\begin{equation}\label{C0}
C(0) = \delta_\nu-\delta_0-\delta~.
\end{equation}
To order $x$ we have one equation determining $b_{0,1}$ and an other
equation determining, using the obtained $b_{0,1}$, $C'(0)$; to order
$x^2$ two equations determining $b_{0,2},b_{1,1}$ and a third equation
determining $C''(0)$ and so on.  At order $x^N$ 
the result of the expansion in $x$ can be written as a finite sum
of terms
\begin{equation}
\frac{1}{(v-1)v^k}~.
\end{equation}  
The equation determining the column vector $(b_{0,N},\dots b_{N-1,1})_t$ is
\begin{equation}\label{AbVequation}
  A^{(N)}
  \begin{pmatrix}
    b_{0,N}\\
    \dots \\
    b_{N-1,1}
  \end{pmatrix}
  = V^{(N)}
\end{equation}  
where $V^{(N)}$ is provided by the results of the previous orders and
the matrix $A^{(N)}_{jk}$
is given by the coefficient of
\begin{equation}
\frac{x^N}{(v-1)v^{2+j}} b_{k-1,N-k+1} ~.
\end{equation}
The vector $V^{(N)}$ depends linearly on the $C^{(n)}$ with $n<N$,
which have been already computed, and depends non linearly from the
$b_{j,k}$ with $j+k<N$ which also have been computed in the previous
steps.

From the structure of the ${\cal B}_j$ we see that the $A^{(N)}_{jk}$
vanish for $k>N$. From the expansion in $x$ of the $Q_v(v)$
eq.(\ref{Qv}) we have that also for $j>N$, $A^{(N)}_{jk}$
vanishes. Again from the structure of the ${\cal B}_j$ we see that for
$j\leq N,k\leq N$ the $A^{(N)}_{jk}$ do not depend on $N$.  Thus the
matrices $A^{(N)}$ are nested matrices. Finally from the fact that the
$b_{j,k}$ appear in the combination ${\cal B}_j/v^j$ and the expansion 
of (\ref{Qv}) we have that $A^{(N)}_{jk}$ vanishes for
$j>k$. Summarizing the matrices $A^{(N)}$ are nested upper triangular
matrices and thus it is sufficient to provide the $N$-th column of the
matrix $A^{(N)}$.

The $N$-th column of $A^{(N)}$ is with $\delta_2 \equiv
\delta_\infty-\delta_1-\delta_\nu$
\begin{equation} \label{Acolumn}
  \begin{pmatrix}
    -4\delta_1 -\delta_2\\
    -6\delta_1 -\delta_2\\
    -8\delta_1 -\delta_2\\
    .... \\
    -2(n+1)\delta_1 -\delta_2\\
.... \\
     -2(N-1)\delta_1 -\delta_2\\
   2 N \delta_\nu+ N(N^2-1)/2+(2N-1)\delta_2\\
 -2 N \delta_\nu- N(N^2-1)/2
\end{pmatrix}
\end{equation}
where $n$ is the row index, and from this due to the nested nature of
the matrices, the whole matrix $A^{(N)}$ can be reconstructed.  Thus
from the third column included the column starts with
$-4\delta_1-\delta_2$ i.e.  $A^{(N)}_{1,N} = -4\delta_1-\delta_2$ for
$N \geq 3$.  For clearness sake we give the $A^{(4)}$ explicitly

\begin{equation}
  \begin{pmatrix}
    -2\delta_\nu & 3 + 4\delta_\nu+3 \delta_2 &-4\delta_1-\delta_2
    &-4\delta_1-\delta_2\\
    0&-3-4\delta_\nu&12+6\delta_\nu+5\delta_2&
    -6 \delta_1 - \delta_2\\
    0&0&-12-6\delta_\nu&30+8\delta_\nu+7\delta_2\\
    0&0&0&-30-8\delta_\nu
\end{pmatrix}~.
\end{equation}

An other writing of the matrix is  
\begin{equation}
  A^{(4)}=\begin{pmatrix}
    -2\delta_\nu & 3(1-\delta_1+\delta_\infty)+\delta_\nu&-3\delta_1+\delta_\nu-
    \delta_\infty
    &-3\delta_1+\delta_\nu-\delta_\infty\\
    0&-3-4\delta_\nu&12-5\delta_1+\delta_\nu+5\delta_\infty&
    -5\delta_1+\delta_\nu-\delta_\infty\\
    0&0&-12-6\delta_\nu &30+\delta_\nu+7\delta_\infty-7\delta_1\\
    0&0&0&-30-8\delta_\nu
\end{pmatrix}~.
\end{equation}

We remark that in the computation of the Schwarz derivative the
denominator in (\ref{ztovchange}) does not contribute and that the
Schwarz derivative contributes only through the numerical terms which
are not multiplied by the $\delta$'s in the last two rows of
eq.(\ref{Acolumn}).  The vanishing of the coefficient of
$\frac{x^N}{v^2(v-1)}$ provides the value of $C^{(N)}$ which is given
by
\begin{equation}
 C^{(N)}/N! = -(w_N,\beta_N)
\end{equation}  
with
\begin{equation}
w_N = (\delta_\nu+\delta_\infty-\delta_1, \delta_\nu-\delta_\infty-\delta_1, 
\dots, \delta_\nu-\delta_\infty-\delta_1)
= (2\delta_\nu+\delta_2, -2 \delta_1-\delta_2 
\dots,-2 \delta_1-\delta_2)
\end{equation}
where all terms except the first one are equal to $-2
\delta_1-\delta_2$ and $\beta_N$ is the column vector
$\beta_N=(b_{0,N},b_{1,N-1},\dots,b_{N-1,1})_t$.

The results one obtains for $C'(0),C''(0),C'''(0)$
\cite{sphere4,sphere5} are collected in Appendix I and agree with the
classical limit of the results available in the literature
\cite{MMM,LLNZ,ferraripiatek}.

We notice that the only operation which intervenes in the
computational process is the inversion of the matrix $A^{(N)}$ which
is also upper triangular.  However such an inversion does not require
the computation of determinants and minors. In fact the inverse of the
matrix
\begin{equation}
  \begin{pmatrix}
     A^{(N)} & r \\ 
     0 & c
  \end{pmatrix}
\end{equation}
where $r,c$ are given by the column $N+1$ is
\begin{equation}
\begin{pmatrix}
     A^{(N)-1} & - A^{(N)-1}~r/c \\ 
     0 & 1/c
  \end{pmatrix}
\end{equation}
and $A^{(N)-1}$ is provided explicitly by the previous step and thus
the only operation required is the multiplication $A^{(N)-1}~r/c$.

We notice that the l.h.s. of eq.(\ref{Qequation}) contributes to the
vector $V^{(N)}$ with the term

\begin{equation}
  \begin{pmatrix}
     -C^{(N-1)}/(N-1)!\\ 
     \dots \\
    - C''(0)/2!\\
    (N+1)\delta- C'(0)\\
    -(N+1)\delta-C(0)
  \end{pmatrix}
\end{equation}
where the $C^{(n)}$ are known from the previous steps 
and $\delta$ appears only in the last two rows. The remainder of the
vector $V^{(N)}$ is provided by terms non linear in the $b_{j,k}$
arising by the expansion of the r.h.s. of eq.(\ref{Qequation}).

\bigskip

A different though equivalent procedure is obtained by equating the
coefficients of $1/v^j$ in the expansion of (\ref{Qequation}).

The column (\ref{Acolumn}) goes over to
\begin{equation}\label{hatAcolumn}
  \begin{pmatrix}
    -(N+1)(N-2)\delta_1 + (N+1)\delta_2\\
    \dots~~~~~~~~~~~~\\
    -(N+n)(N-n-1)\delta_1+(N+n)\delta_2\\
    \dots            \\
     ~~~~~~~~~~-2(N-1)\delta_1+(2N-2)\delta_2\\ 
  ~~~~~~~~~~~~~~~~~~~~~~~~~~~~~~~~(2N-1)\delta_2\\
 -2 N \delta_\nu- N(N^2-1)/2 
\end{pmatrix}
\end{equation}
giving rise to a new matrix $\hat A^{(N)}$.

This has the advantage over $A^{(N)}$ that the $O(N^3)$ dependence
is confined to the last row of (\ref{hatAcolumn}).

The transformation which takes $A$ into $\hat A$ is $\hat A= S A$ where

\begin{equation}
  S=
  \begin{pmatrix}
    1 & 1 & 1 & \dots 1\\
    0 & 1 & 1 & \dots 1\\
    0 & \dots & 1 &\dots 1\\
    0 & 0 & 0 & \dots 1
\end{pmatrix}~.
\end{equation}
is the upper triangular matrix with all non zero entries equal to $1$.
The vector which appears on the r.h.s. is accordingly changed. The
contribution of the r.h.s. of eq.(\ref{Qequation}) to it now becomes
\begin{equation}
  \begin{pmatrix}
    - C(0)- C'(0)-C''(0)/2!-\dots -C^{{(N-1)}}/(N-1)!\\
     \dots \\ 
     -C(0)-C'(0) \\
     (N+1)\delta - C(0)
\end{pmatrix}
\end{equation}
and also the non linear contributions due to the lower order $b_{j,k}$
are changed but not the $b_{j,k}$'s themselves.

The described algebraic procedure allows to compute in a
straightforward way the series in $x$ for the $C(x)$
to all orders.
What is left open is the nature of this series i.e. if it is simply a
formal power series, an asymptotic series or a convergent series
giving rise to an analytic function.

We saw how the key role is played by the inverse of the matrices
$A^{(N)}$.
One can give rigorous bounds on the norms of the matrices 
$(A^{(N+1)})^{-1}$.
A bound which holds both in the norm ${\rm Sup}_k |v_k|$ and in the
$l^2$ norm is obtained  starting from
\begin{equation}
  \begin{pmatrix}
     A^{-1} &- A^{-1} \frac{r}{c} \\ 
     0 & \frac{1}{c}
  \end{pmatrix}
  \begin{pmatrix}
    v \\
    a
  \end{pmatrix}
\end{equation}  
giving rise to
\begin{equation}
  ||(A^{(N+1)})^{-1}||\leq
  ||(A^{(N)})^{-1}||\bigg(1+\frac{||r||}{|c|}\bigg)+\frac{1}{|c|}~.
\end{equation}  
A better bound is obtained for the matrices $\hat A$ 
where it gives for large $N$ the bound in the ${\rm Sup}_k |v_k|$ norm
\begin{equation}
  ||(\hat A^{(N+1)})^{-1}||\leq
  ||(\hat A^{(N)})^{-1}||~\big(1+\frac{2 |\delta_1|}{N}\big)+\frac{1}{N^3}
\end{equation}  
i.e. a bound which behaves for large $N$ like $N^{2|\delta_1|}$.
However from the purely algebraic viewpoint while the $C^{(n)}$,
$n<N$ contribute linearly to the vector $V^{(N)}$ it is very difficult
to control the non linear contributions of the $b_{j,k}$, $j+k<N$,
to $V^{(N)}$.

The problem of the convergence of the series is addressed in the
following section.

\bigskip

\section{The Green function approach}\label{green}

In this section we give a non perturbative computation of the
monodromy around the contour II described in section \ref{algebraic}.
The procedure is
rigorous and as a by product we shall prove the convergence of the
power series for $C(x)$ derived in section \ref{algebraic} with a
finite (non zero) convergence radius.

We shall consider only elliptic singularities and monodromies i.e
according to the usual classification, monodromies
with the square of the trace positive and less than $4$.
We have for all $\delta_j$
\begin{equation}
\delta_j=\frac{1-\lambda^2_j}{4}
\end{equation}  
and for the associated monodromy we have ${\rm tr} M_j= -2
\cos(\pi\lambda_j)$. The monodromy is unchanged for
$\lambda\rightarrow -\lambda$ and for $\lambda\rightarrow \lambda+2
n$.  We shall choose by definition $0<\lambda_j$. All real traces with
square less than $4$, i.e. the traces of all elliptic monodromies, can
be obtained with $\lambda_j<1$ and we shall work in this region. This
corresponds to having all $\delta$'s, $\delta_\nu$ included, positive
and less than $1$.

After writing $Q(z)=Q_0(z)+\Delta(z)$ the iterative scheme for
\begin{equation}\label{eqiterative}
  y''(z)+ Q_0(z) y(z) = -\Delta(z)y(z),
\end{equation}
where 
\begin{equation}
  \Delta(z)=x \frac{(\delta_0-\delta_\nu)(z-x)+\delta~z(2z-x-1)}
         {z^2(z-x)^2(z-1)}-
         \frac{c}{z(z-x)(z-1)}
\end{equation}
with
\begin{equation}
  c= C-C(0)~,
\end{equation}
gives rise for the solution $y^G_j$ of (\ref{eqiterative}) to 
\begin{equation}
  y^G_j(z) = y_j(z)+\int_1^z g(z,z')\Delta(z')y_j(z')dz'
  +\int_1^z g(z,z')dz'\Delta(z')g(z',z'')\Delta(z'')y_j(z'')dz''+\dots
\end{equation}
The Green function $g(z,z')$ is 
\begin{equation}
  g(z,z') = \frac{1}{w_{12}} \bigg(-y_1(z) y_2(z')+
  y_2(z) y_1(z')\bigg) \Theta(z,z')
\end{equation}
with $w_{12}= y_1'(z) y_2(z)-y_1(z) y'_2(z)={\rm const}$.
$\Theta(z,z')$ is the step function evaluated along the chosen
integration path, and $y_1$, $y_2$ are two independent solutions
of $y''+Q_0y=0$.  We notice that $\Delta$ as a function of $z$
is not singular in the integration range $(1,\infty)$.  The values
of two independent solutions above the cut in $z$ running from $1$ to
$+\infty$ are

\begin{eqnarray}\label{hypergeometric1}
y^+_1(z)&=&(1-z)^{\frac{1-\lambda_1}{2}}~z^{\frac{1-\lambda_\nu}{2}}
F(\frac{1-\lambda_1-\lambda_\infty-\lambda_\nu}{2},
\frac{1-\lambda_1+\lambda_\infty-\lambda_\nu}{2},1-\lambda_1;1-z)
\nonumber\\
&\equiv& -i e^\frac{i\pi\lambda_1}{2}t_1(z)
\end{eqnarray}
\begin{eqnarray}\label{hypergeometric2}
y^+_2(z)&=&(1-z)^{\frac{1+\lambda_1}{2}}~z^{\frac{1+\lambda_\nu}{2}}
F(\frac{1+\lambda_1+\lambda_\infty+\lambda_\nu}{2},
\frac{1+\lambda_1-\lambda_\infty+\lambda_\nu}{2},1+\lambda_1;1-z)\nonumber\\
&\equiv&-i e^{-\frac{i\pi\lambda_1}{2}} t_2(z)~.
\end{eqnarray}

The solutions of eq.(\ref{fundeq}) are
\begin{equation}
(1+S_{11}(z))~ y_1(z) + S_{12}(z) ~y_2(z)\\
\end{equation}
and
\begin{equation} 
S_{21}(z) ~y_1(z)+ (1+S_{22}(z))~ y_2(z)
\end{equation}
where

\begin{eqnarray}\label{S11}
 && -w_{12}S_{11}(z)= \int_1^z y_2(z')\Delta(z')y_1(z')dz'+\\
 && \int_1^z y_2(z')\Delta(z')dz'
  g(z',z'')\Delta(z'')y_1(z'')dz''+\nonumber\\
 &&\int_1^z y_2(z')\Delta(z')dz'
  g(z',z'')\Delta(z'')dz''g(z'',z''')\Delta(z''')y_1(z''')dz'''+\dots
\nonumber
\end{eqnarray}

\begin{eqnarray}\label{S12}
 && w_{12}S_{12}(z)= \int_1^z y_1(z')\Delta(z')y_1(z')dz'+\\
 && \int_1^z y_1(z')\Delta(z')dz'
  g(z',z'')\Delta(z'')y_1(z'')dz''+\nonumber\\
 &&\int_1^z y_1(z')\Delta(z')dz'
  g(z',z'')\Delta(z'')dz''g(z'',z''')\Delta(z''')y_1(z''')dz'''+\dots\nonumber
\end{eqnarray}

\begin{eqnarray}\label{S21}
 && -w_{12}S_{21}(z)= \int_1^z y_2(z')\Delta(z')y_2(z')dz'+\\
 && \int_1^z y_2(z')\Delta(z')dz'
  g(z',z'')\Delta(z'')y_2(z'')dz''+\nonumber\\
 &&\int_1^z y_2(z')\Delta(z')dz'
  g(z',z'')\Delta(z'')dz''g(z'',z''')\Delta(z''')y_2(z''')dz'''+\dots
  \nonumber
\end{eqnarray}

\begin{eqnarray}\label{S22}
 && w_{12}S_{22}(z)= \int_1^z y_1(z')\Delta(z')y_2(z')dz'+\\
 && \int_1^z y_1(z')\Delta(z')dz'
  g(z',z'')\Delta(z'')y_2(z'')dz''+\nonumber\\
 &&\int_1^z y_1(z')\Delta(z')dz'
  g(z',z'')\Delta(z'')dz''g(z'',z"'')\Delta(z''')y_2(z''')dz'''+\dots
  \nonumber
\end{eqnarray}

In order to examine the convergence of the series it is useful
to go over to the following quantities denoted by a hat

\begin{equation}
  y_k(z) = (1-\frac{1}{z})^{\frac{1-\lambda_1}{2}}z^{\frac{1+\lambda_\infty}{2}}
  \hat y_k(z)
\end{equation}

\begin{equation}
  g(z',z'')= (1-\frac{1}{z'})^{\frac{1-\lambda_1}{2}}
  z'^{\frac{1+\lambda_\infty}{2}}\hat g(z',z'')
  (1-\frac{1}{z''})^{\frac{1-\lambda_1}{2}}
z''^{\frac{1+\lambda_\infty}{2}}
\end{equation}

\begin{equation}
  \hat \Delta(z) =z^{1+\lambda_\infty}(1-\frac{1}{z})^{1-\lambda_1}\times \Delta(z)~. 
\end{equation}
$|\hat g(z,z')|$ can be majorized by a constant $g_M$ (depending on
$\lambda_1,\lambda_\nu,\lambda_\infty$) and thus $S_{11}$, see
eq.(\ref{S11}), can be majorized by

\begin{eqnarray}
&&  \frac{M^2}{w_{12}}\bigg(\int_1^z|\hat\Delta(z')| dz'+
  \frac{g_M}{2}\bigg(\int_1^z|\hat\Delta(z')| dz'\bigg)^2+\dots\bigg)
  \leq\nonumber\\
&&  
  \frac{M^2}{w_{12}}\frac{1}{g_M}
  \bigg(\exp\big(g_M \int_1^z|\hat\Delta(z')| dz'\big) -1\bigg)
\end{eqnarray}
where $M=\max |\hat y_j|$ i.e. the maximum of $|\hat y_j(z)|$ for $1\leq
z<\infty, j=1,2$~.

A very simple bound on $\hat\Delta(z,x,c)$ for $z>1$ is
\begin{eqnarray}
  &&  |\hat \Delta(z,x,c)| \leq
  z^{1+\lambda_\infty}(1-\frac{1}{z})^{1-\lambda_1}\times\\
  && 
  \bigg[|x|\frac{|\delta_0-\delta_\nu|}
  {z^2|z-x|(z-1)}+|x|\delta \bigg(\frac{1}{z|z-x|^2}+\frac{1}{z(z-1)|z-x|}\bigg)
 +\frac{|c|}{z(z-1)|z-x|}\bigg]~. \nonumber
\end{eqnarray}

The same bound holds for the other $S_{jk}(z)$.
As the integral
\begin{equation}
\int_1^\infty|\hat\Delta(z',x,c)| dz'
\end{equation}
is finite for $|x|<1$ and any $c$, the series for the solution
converges for $1<z<\infty$ and as such it is a non perturbative
result.  The convergence is also uniform. We notice that this holds
also for $z=+\infty$, which is the value which intervenes in the
expression of the monodromy.

We remark that despite the expansion of the $Q(z,x)$ is very singular
for $x=0$ if we work, as we do, along the line $z$ with
$1\leq z\leq\infty$ the
perturbation is a regular perturbation.

Then the monodromy along the cycle II of fig.1  changes to

\begin{equation}
  M=
  \begin{pmatrix}
      1+S^+_{11} &  S^+_{12}\\
      S^+_{21}& 1+S^+_{22}
  \end{pmatrix}
  M^0
  \begin{pmatrix}
      1+S^-_{22} &  -S^-_{12}\\
      -S^-_{21}&1+S^-_{11}
  \end{pmatrix}
\end{equation}
where
\begin{equation}
  M^0= - B^+
  \begin{pmatrix}
      e^{-i\pi\lambda_\infty}&  0\\
      0 &e^{i\pi\lambda_\infty}
  \end{pmatrix}
  (B^-)^{-1}
\end{equation}
and the $S^{\pm}_{jk}$ denote $S^{\pm}_{jk}(+\infty)$ with the integral
computed above and below the cut $(1,\infty)$.

The matrix
\begin{equation}
    \begin{pmatrix}
      1+S^+_{11}(+\infty) &  S^+_{12}(+\infty)\\
      S^+_{21}(+\infty) & 1+S^+_{22}(+\infty)
  \end{pmatrix}
\end{equation}
is $SL(2C)$ due to the constancy of the Wronskian.

The explicit form of $B$ is
\begin{equation}\label{Bmatrix}
B=
\begin{pmatrix}
\frac{\Gamma(1-\lambda_1) \Gamma(-\lambda_\infty)}
{\Gamma\big(\frac{1-\lambda_1-\lambda_\infty-\lambda_\nu)}{2}\big)
\Gamma\big(\frac{1-\lambda_1-\lambda_\infty+\lambda_\nu}{2}\big)}&
\frac{\Gamma(1-\lambda_1) \Gamma(\lambda_\infty)}
{\Gamma\big(\frac{1-\lambda_1+\lambda_\infty-\lambda_\nu}{2}\big)
\Gamma\big(\frac{1-\lambda_1+\lambda_\infty+\lambda_\nu}{2}\big)}\\
\frac{\Gamma(1+\lambda_1) \Gamma(-\lambda_\infty)}
{\Gamma\big(\frac{1+\lambda_1-\lambda_\infty+\lambda_\nu)}{2}\big)
\Gamma\big(\frac{1+\lambda_1-\lambda_\infty-\lambda_\nu}{2}\big)}&
\frac{\Gamma(1+\lambda_1) \Gamma(\lambda_\infty)}
{\Gamma\big(\frac{1+\lambda_1+\lambda_\infty+\lambda_\nu}{2}\big)
\Gamma\big(\frac{1+\lambda_1+\lambda_\infty-\lambda_\nu}{2}\big)}
\end{pmatrix}
\end{equation}
while $B^+$ and $B^-$ are given by eq.(\ref{Bplusminus}).
\bigskip
Then we have
\begin{equation}\label{traceequation}
  {\rm tr} (M-M^0)=
  {\rm tr}(S^+M^0)+{\rm tr}(M^0 \tilde S^-)+
  {\rm tr}(S^+ M^0 \tilde S^-)
\end{equation}
where $\tilde S$ is defined as
\begin{equation}
  \tilde S =
  \begin{pmatrix}
      S_{22} &  -S_{12}\\
      -S_{21}& S_{11}
  \end{pmatrix}~.
\end{equation}
We notice that for $1\leq z<\infty$, $\hat\Delta$
as a function of the two variables $x$ and $c$ is analytic in the
polydisk given by $|x|<\tau$ with any $0<\tau<1$ and any $c$
and given $x$ we have to find a $c$ such that
\begin{equation}
  {\rm tr} (M-M^0)=0~.
\end{equation}

To first order in $x$ we have
\begin{equation}\label{firstorder}
{\rm tr}~ M = {\rm tr}~ M^0 +x~{\rm tr}~[({\cal S}^+-{\cal S}^-) M^0 ] ~~.
\end{equation}
with
\begin{equation}
  {\cal S} =\frac{1}{w_{12}}
    \begin{pmatrix}
      -\int_1^\infty y_2 Q_1 y_1 dz & \int_1^\infty y_1 Q_1 y_1 dz \\
       -\int_1^\infty y_2 Q_1 y_2 dz & \int_1^\infty y_1 Q_1 y_2 dz 
\end{pmatrix}  
\end{equation}

\begin{eqnarray}
  && y^+_1 = -i e^\frac{i\pi\lambda_1}{2}~t_1\\
  && y^+_2 = -i e^{-\frac{i\pi\lambda_1}{2}} ~t_2
\end{eqnarray}  

\begin{eqnarray}
  && y^-_1 = i e^{-\frac{i\pi\lambda_1}{2}}~t_1\\
  && y^-_2 = i e^{\frac{i\pi\lambda_1}{2}} ~t_2~.
\end{eqnarray}  
We have
\begin{equation}
({\cal S}^+-{\cal S}^-)_{11}=-({\cal S}^+-{\cal S}^-)_{22}=0
\end{equation}  
and thus
\begin{equation}
  {\rm tr} (\delta M) = x
  \bigg[({\cal S}^+-{\cal S}^-)_{12} M^0_{21}
    +({\cal S}^+-{\cal S}^-)_{21}M^0_{12}\bigg]=0 ~.
\end{equation}

\bigskip

For the values of $M^0_{12},M^0_{21}$ we have
\begin{equation}
  M^0\times {\rm det}(B) =\frac{1}{w_{12}}
    \begin{pmatrix}
      * & 2 i \sin\pi\lambda_\infty B_{11}B_{12}\\
      -2 i \sin\pi\lambda_\infty B_{21}B_{22} & *
\end{pmatrix} ~. 
\end{equation}

But 
\begin{eqnarray}
&&({\cal S}^+-{\cal S}^-)_{12}= \frac{2i \sin\pi\lambda_1}{w_{12}}Q_1(1,1)\\
&&({\cal S}^+-{\cal S}^-)_{21}= \frac{2i \sin\pi\lambda_1}{w_{12}}Q_1(2,2)
\end{eqnarray}
with
\begin{equation}
Q_1(j,k)=\int_1^\infty t_j Q_1 t_k dz
\end{equation}
where
\begin{equation}
  Q_1 = \frac{2\delta - C'(0)}{z^2(z-1)}-
  \frac{2\delta + C(0)}{z^3(z-1)}.
\end{equation}
Thus the first order equation becomes
\begin{equation}\label{Q1122equation}
B_{21} B_{22} Q_1(1,1)-B_{11} B_{12} Q_1(2,2)=0~.
\end{equation}
To compute the previous
we need integrals of the type
\begin{equation}  
\int_1^\infty t_j\frac{1}{z^m(z-1)} t_k dz~.
\end{equation}
In Appendix 2 we give the general technique to compute such integrals.
For the first order computation we shall need only
the values for $m=2$ and $m=3$.
To compute (\ref{Q1122equation}) let us perform on the
\begin{equation}
  Q_0 = \frac{\delta_\nu}{z^2}+\frac{\delta_1}{(z-1)^2}
  +\frac{\delta_\infty-\delta_1-\delta_\nu}{z(z-1)}
\end{equation}
the infinitesimal dilatation with center $z=1$ i.e.
\begin{equation}
z = \frac{v-\varepsilon}{1-\varepsilon}~.
\end{equation}
The change in $Q_0$
\begin{equation}  
\tilde Q_0(v) = Q_0(z(v)) (z'(v))^2-\{z,v\}~,
\end{equation}   
as now $\{z,v\}=0$ is given by $\varepsilon R_1$ where 
\begin{equation}  
R_1=\frac{\delta_\infty+\delta_\nu-\delta_1}{z^2(z-1)}
-\frac{2\delta_\nu}{z^3(z-1)}~.
\end{equation}
Then we can write
\begin{equation}
  Q_1 = \frac{1}{z^2(z-1)}
  \bigg(2\delta-C'(0) -\frac{(2\delta+C(0))(\delta_\infty-
    \delta_1+\delta_\nu)}{2\delta_\nu}\bigg)
+\frac{2\delta+C(0)}{2\delta_\nu} R_1~.
\end{equation}
From Appendix 2 we have
\begin{eqnarray}\label{R1elements}
&& \int_1^\infty t_1(z) R_1(z) t_1(z) = -\lambda_\infty B_{11} B_{12}\nonumber\\
&&\int_1^\infty t_2(z) R_1(z) t_2(z) = -\lambda_\infty B_{21} B_{22} ~.
\end{eqnarray}
From eq.(\ref{R1elements}) we see that the
contribution of $R_1$ to eq.({\ref{Q1122equation}) is zero. This
  result is expected as the dilatation around $z=1$ does not alter the
  monodromy at infinity. Thus the final result is
\begin{equation}\label{Cprime}
C'(0) = 2\delta -\frac{(2\delta
  +C(0))(\delta_\infty-\delta_1+\delta_\nu)}{2\delta_\nu}~.
\end{equation}

In \cite{sphere4} the explicit calculation of the second order in $x$
was given and the result is reported in Appendix 1. Both results agree
with the $C'(0)$ and $C''(0)$ obtained in the algebraic approach.
 
\bigskip

In complete generality we can state that the trace of the monodromy
$M_{0x}$ is an analytic function of the two variable in the polydisk
$(|c|<R) \times(|x|<\tau)$ for any $0<\tau<1$ and any $R$. The problem posed in
the introduction is, given $x$, to find a $c$ such that the trace of
such a monodromy equals $-2\cos\pi\lambda_\nu$. From the previous
computation we have found that at $x=0$ the derivative of such a trace
w.r.t. $c$ is different from zero. Then from the implicit analytic
function theorem it follow that in a finite but non zero disk
$|x|<\varepsilon$ the above equation has a unique solution $c(x)$ and
such solution is analytic in $x$.

We saw by explicit calculation that the first and second derivative of
$c(x)$ computed in the algebraic approach coincide with the present
implicit function approach. But this holds to all orders.

In fact we have proven in the Green function approach that for
$|x|<\varepsilon$ there exists one and only one $C(x)$ such the
monodromy $M_{0x}$ has trace $-2\cos \pi\lambda_\nu$.  Thus given an
order $n$ we can expand $y_j$ up to order $n$ and these function
satisfy the equation (\ref{fundeq}) up to order $n$ included.  By the
way, for the computation of such polynomial only the first $n$ terms
of the series (\ref{S11}-\ref{S22}) are necessary.

In the Green function approach we have to all orders for $z$ near $1$
\begin{equation}
  y_1^G(z)\approx(z-1)^{\frac{1-\lambda_1}{2}},~~~~~~~~
  y_2^G(z)\approx(z-1)^{\frac{1+\lambda_1}{2}}  
\end{equation}}
and we proved that $y_1,y_2$ are analytic functions of $x$, $|x|<1$
and any $c$ and that for $|x|<\varepsilon$ there exists only one
choice of $c$, $c(x)$ for which the two $y_j$ give rise to the
prescribed monodromy and that $c(x)$ is analytic in $x$.

In the algebraic treatment of section (\ref{algebraic}) we started
from a parametrization of the solutions i.e.
\begin{equation}
  y^A_j(v) = y_j(z(v)) (z'(v))^{-\frac{1}{2}}
\end{equation}  
which due to $v\sim z$ for $v\rightarrow \infty$ has the trace of the
monodromy $M_{0,x}$ unchanged i.e. equal to $-2\cos \pi\lambda_\nu$
and we fix order by order in $x$ the parameters $b_{j,k}$ and
$C(0),C(0)', C(0)'' \dots $ as to have those function to satisfy the
equation
\begin{equation}
{y^A}''(v)+ Q(v) y^A(v)=0
\end{equation}
up to order  $n$ included. We have at $v=1$
\begin{equation}
  y_1^A(v)\sim (v-1)^{\frac{1-\lambda_1}{2}},~~~~
  y_2^A(v)\sim (v-1)^{\frac{1+\lambda_1}{2}}~.
\end{equation}

Thus, apart for a multiplicative constant, the functions $y_j^G$ and
$y_j^A$ coincide up to order $n$ and the coefficients of the expansion
of $C$ in the algebraic expansion coincide with the expansion of the
analytic function $C(x)$ found in Green function approach.

The only thing that was missing in the algebraic approach was the
proof of the convergence of the series which has been given for
$|x|<\varepsilon$ in the present section.

More difficult is to provide a rigorous lower bound to the convergence
radius of the power series in $x$.  One can follow the procedure of
constructing a ``good polydisk'' in Whitney terminology
\cite{whitney}.  A good polydisk in $c$ and $x$ for the function
$h(x,c)$ is defined as a polydisk $P=\Delta_x\times \Delta_c$ with
$\Delta_x$ given by $|x|<b_x$ and $\Delta_c$ given by $|c|<b_c$ such
that $h(x,c)$ is holomorphic in $P$ and $h(x,c)$ does not vanish on
$\Delta_x\times\partial\Delta_c$.  On the whole polydisk $P$ the
$h(x,c)$ is then represented by a Weierstrass polynomial. Moreover as
we shall see below, we can choose $P$ so that $h(0,c)=0$ has a single
simple solution in $P$ and $h(x,c)\neq 0$ on $\Delta_x\times\partial
\Delta_c$ i.e. $h(x,c)\neq 0$ for $(x,c)\in\Delta_x \times\{c:|c|=b_c\}$. This
from the Weierstrass polynomial representation implies the analyticity
of the solution in the whole polydisk.

A method to construct such a good polydisk is to explicitly sum
the first $n$ terms of the series giving $h(x,c)$ and to put a bound
on the remainder of the series.

We illustrate below the procedure in the simplest, even though not
very efficient, choice where one writes
\begin{equation}\label{hbreakdown} 
{\rm tr}(M-M^0)\equiv h(x,c)=c f_1(x)+g_1(x) + r(x,c)
\end{equation}
where $c f_1(x)+g_1(x)$ is the first order in $\Delta$ of
eq.(\ref{traceequation}) i.e.
\begin{equation} 
  c f_1(x)+g_1(x) = M^0_{12}(S^+_{1,21}-S^-_{1,21})+
                    M^0_{21}(S^+_{1,12}-S^-_{1,12})
\end{equation}
where
\begin{eqnarray} 
  && S_{1,12}=\frac{1}{w_{12}}\int_1^\infty y_1(z')\Delta(z')y_1(z') d
  z'\nonumber\\ && S_{1,21}=-\frac{1}{w_{12}}\int_1^\infty
  y_2(z')\Delta(z')y_2(z') d z'
\end{eqnarray}
which is a known function and $r(x,c)$ is the remainder.
We notice that
\begin{equation}
f_1(0)\neq 0 ,~~~~g_1(0)=0~.
\end{equation}

Straightforward inequalities provide
\begin{equation}\label{rbound}
  |r(x,c)|< m_0\big[2(e^\eta-1-\eta)+(e^\eta-1)^2\big]
\end{equation}
where $m_0 = \max(|M^0_{i,1}|+|M^0_{i,2}|)$ and
\begin{equation}
  \eta = 2~ \max_j\int_1^\infty \frac{\hat t^2_j(z)}{w_{12}}|\hat\Delta(z)| dz
\equiv |c|F(x)+ G(x)~.
\end{equation}

If we denote by $c_1$ the positive root of the equation
\begin{equation}
|f_1(0)|c_1 -m_0\big(2(e^{c_1F(0)}-1-c_1F(0)+ (e^{c_1F(0)}-1)^2\big)=0
\end{equation}
we have for $0<|c|<c_1$
\begin{equation}
f_1(0)c+ r(0,c)\neq 0~.
\end{equation}
Then for every $c$, with $|c|=s$ and $s$ in the interval
 $(0,c_1)$ we have a disk $\Delta_x$, $|x|<\beta(s)$
in which
\begin{equation}
  |c| |f_1(x)|-|g_1(x)| - m_0(e^{|c|F(x)+G(x)}-1-|c|F(x)-G(x)+
  (e^{|c|F(x)+G(x)}-1)^2)>0~.
\end{equation}

Then on $\partial\Delta_c\times\Delta_x$, where $\Delta_c$ is the disk
$|c|<s$  we have
\begin{equation}
|h(x,c)| > 0 
\end{equation}
which is the definition of a good polydisk.
One has to choose $s$ in the range $(0,c_1)$ so that
the $\beta(c)$ is the largest.

As an illustration applying the described procedure to the case when
all $\lambda$'s equal $0.5$, we obtain a rigorous lower bound of
0.0175 for the convergence radius of the power expansion in $x$.  The
smallness of such lower bound is due to the very naive decomposition
(\ref{hbreakdown}). A larger convergence radius may be provided by
computing explicitly more terms of the series for $h(x,c)$ and putting
a bound on the new remnant, a rather laborious task that we won't
pursue here. On the other hand it could well happen that the some
singularities of the accessory parameters are present near zero.

From the general viewpoint the rigorous results of the present section
tell us that provided one ascertain the existence and uniqueness of
the $x$ solving the equation ${\rm tr}(M-M^0)\equiv h(x,c)=0$ then we
have analyticity for $|x|<1$ and thus a convergence radius of $1$. In
fact if we denote by $x_0,c_0$ a solution of
$h(x_0,c_0)=0$ uniqueness tells us that $h(x_0,c)$ around $c_0$
depends on $c$ and thus Weierstrass preparation theorem applies,
i.e. we have a polydisk containing the solution $c_0=c(x_0)$ where
\begin{eqnarray}\label{uniquesol}
  &&  h(x,c)= u(x,c)(c-c(x))^N = \\
  && u(x,c)\bigg((c-c_0)^N + N (c-c_0)^{N-1}(c_0-c(x))+
  \dots + (c_0-c(x))^N\bigg)\nonumber
\end{eqnarray}
where $u(x,c)$ is a unit, $N$ is the order of $h$ at $x_0,c_0$ and due
to the analyticity of the coefficients of the Weierstrass polynomial,
in particular the coefficient of $(c-c_0)^{N-1}$, we have that $c(x)$
analytic. This holds for all $|x|<1$ where, due to the proved
convergence of the series, $h(x,c)$ is analytic.

For the accessory parameters $C(x,\bar x)$ related to the
uniformization problem  we described in section \ref{classicallimit},
we have constructive proofs
\cite{poincare,lichtenstein,menottisolution} of the existence and
uniqueness of the conformal factor and thus of the accessory
parameters, in addition to their real analytic dependence on the
position of the singularities
(\cite{menottielliptic,menottiparabolic}).

On the other hand we are not aware of a proof of existences and/or
uniqueness of the accessory parameters related to the problem of
\cite{LLNZ} when $|x|$ is not small. For small $|x|$ the existence and
uniqueness proof was given in the present section.

\section{Relation to the Riemann-Hilbert problem}\label{riemannhilbert}

There is a strong similarity between the problem of \cite{LLNZ} of
proving the existence and the uniqueness of the accessory parameter
which induces ${\rm tr} M_{(0,x)} = -2\cos\pi \lambda_\nu$ and the
Riemann-Hilbert (R-H) problem i.e. the 21st Hilbert problem.

The R-H problem i.e. to find a fuchsian ODE which gives rise to a
complete set of given monodromies up to a common similitude
transformation, has been completely understood \cite{bolibrukh}. For
monodromies of dimension two, which are related to a second order
differential equations the R-H problem is always soluble given any set
of monodromies \cite{bolibrukh,deligne,dekkers}. In particular 
for second order ODE Dekkers \cite{dekkers} gave a purely algebraic
treatment of the problem.

A general analysis \cite{NRS} shows that the complex dimension of the
$SL(2,C)$ flat connection with four singular points is six, four
of which are given by the
trace around the four singular points, in our case $0,x,1,\infty$.
As local coordinates for these flat connections one can use \cite{NRS}
$m_j={\rm tr}~g_j$ with $i=1,2,3,4$ where the $g_j$ are related by
$g_1g_2g_3g^{-1}_4=1$ and $m_{12}={\rm tr}(g_1 g_2), $ $m_{23}=
{\rm tr}(g_2 g_3)$. Not all the other traces of the product of two and three
$g_j$ are globally expressible in terms of the written traces. The
reason is that \cite{NRS} $m_{13}={\rm tr}g_1 g_3$ is expressible in
terms of the previous through a quadratic relation which make the
above written traces local but not global coordinates.

On the other hand the complex degrees of freedom of ODE taking into
account the Fuchs conditions is $n^2(m+2 g-2)/2+nm/2$ where $n$ is the
order of the equation i.e. the dimension of the representation, and
$m$ the number of singularities and $g$ is the genus. Thus for $n=2$,
$g=0$ and $m =4$ we have $8$ complex degrees of freedom against $9$
degrees of freedom in the general monodromies.

Obviously given the four monodromies whose product equals the
identity, also the traces of the monodromies are given. The
solubility of the R-H problem assures us that given any monodromy
representation of the fundamental group this can be realized by a
fuchsian differential equation.

For $SL(2,C)$ monodromies we have, as said, $6$
degrees of freedom while in eq.(\ref{accessoryeq}) which should
realize them only $5$ i.e. the four $\delta$'s and the accessory
parameter.

Thus we can hope to impose the value of ${\rm tr} M_{0x}=m_{12}$ but
not at the same time the value of ${\rm tr} M_{01}=m_{23}$. In
fact in section \ref{green} we rigorously proved that for small $|x|$
we can impose the value of ${\rm tr} M_{0x}=m_{12}$ and this can be
done in a unique way; the value of ${\rm tr} M_{01}=m_{23}$ becomes
fixed consequently.

On the the other hand in the R-H problem we can freely give in
addition to $\delta_\nu$ also the trace of the loop $(01)$.

This can be seen as follows: referring always to the elliptic
monodromy case, in the R-H setting we can start from the monodromy
around $0$ given by
\begin{equation}
  \begin{pmatrix}
    -e^{-i\pi\lambda_0} &0\\
    0 & -e^{i\pi\lambda_0}
  \end{pmatrix}  
\end{equation}  
with trace $-2\cos \pi \lambda_0$ and with the monodromy around $x$
\begin{equation}
  \begin{pmatrix}
    -e^{-i\pi\lambda} &0\\
    0 & -e^{i\pi\lambda}
  \end{pmatrix}~.  
\end{equation}
Given two monodromies not proportional to the identity with
traces $-2\cos \pi\lambda_0$, $-2\cos \pi\lambda$ is possible to find
a member of the class of the second such that multiplied by the first
gives a member of the class $-2\cos\pi\lambda_\nu$.
In fact the equation 
\begin{equation}\label{obtaindeltanu}
  {\rm tr} M_0 A M_xA^{-1}= 2 \cos \pi (\lambda_0+\lambda)+ 2 c b
  (\cos\pi(\lambda_0+\lambda))- \cos\pi(\lambda_0-\lambda)) = -2\cos
  \pi\lambda_\nu
\end{equation}
with
\begin{equation}
A=  \begin{pmatrix}
    a&b\\
    c&d
\end{pmatrix},
~~~~ad-bc=1    
\end{equation}
can always be solved in the product $c b$. Once this product is
fixed and thus also $a d=cb+1$ is fixed we still have two complex
degrees of freedom which leaves the result unchanged
i.e. multiplication of $c$ by a factor and the same for $a$.
We take the monodromy around $1$ as
\begin{equation}
  \begin{pmatrix}
    -e^{-i\pi\lambda_1} &0\\
    0 & -e^{i\pi\lambda_1}
  \end{pmatrix}~.  
\end{equation}
Thus now the monodromy at infinity is fixed by $M_\infty=
(M_0M_xM_1)^{-1}$.  Adopting the procedure described above we can now
perform a similitude transformation on $M_x$ and $M_1$ such that
\begin{equation}
{\rm tr} M_0 M'_x M'_1 = -2 \cos\pi\lambda_\infty 
\end{equation}
thus realizing the data of our original problem. However all this has
in general the price of the occurrence of apparent (or false)
singularities \cite{bolibrukh}.  These can be chosen to be in
number less or equal to \cite{ohtsuki}
\begin{equation}
  1-n(1-g)+\frac{n(n-1)}{2}(m+2 g -2)
\end{equation}
where $m$, the number of singularities, in our case is $4$ the genus
$g$ equals $0$ while $n$, the dimension of the representation is $2$
and thus the number of apparent singularities can be chosen not higher
than $1$.

However we saw how the followed procedure of reducing out original
problem to a R-H problem has still a degree of freedom of 2 complex
parameters and one expects that for a special choice of $M_{01}$ the
apparent singularity, which is also characterized by two parameters i.e.
the position of the singularity $z_A$ and a residue $\mu_A$
\cite{yoshida}, disappears.

Obviously establishing the existence and uniqueness of the solution is
very important because in this case the rigorous results of section
\ref{green} i.e. the analyticity of $h(x,c)$ for $|x|<1$ and any $c$
implies that, such a parameter depends analytically on the $x$ in the
whole disk $|x|<1$ as was pointed out in section \ref{green} after
eq.(\ref{uniquesol}).

\section{Conclusion and open problems}\label{conclusions}

After describing a very efficient iterative method, which works to all
orders, to provide the coefficients of the expansion of the accessory
parameters in the modulus $x$, we proved in section \ref{green} that
such a series is actually convergent in a disk of non zero radius.
Naively from the criterion of the nearest singularity \cite{ZZ} one
would expect such a radius should be equal to $1$ i.e. the distance
from the nearest singularity of the differential equation.

As we pointed out in section \ref{riemannhilbert} such a convergence
radius is not granted for the classical conformal blocks i.e. for the
accessory parameter.

A direct proof of the convergence of the series in the algebraic
approach appears very difficult: Even if one can give a good bound on
the norm of the inverses of the matrices $A^{(N)}$ the occurrence of
non linear contributions in the equations for the $b_{j,k}$ from the
lower orders makes it very hard to put an effective bound on the
series.

Thus we relied on the method of implicit functions proving that
for small $x$ the series converges and outlining also a procedure
to provide a rigorous lower bound to such convergence radius.
        
The problem treated brings a close relationship to the classical
Riemann-Hilbert problem which has been completely solved
\cite{bolibrukh,dekkers} for any given set of monodromies for a second
order differential equation, a problem where more detailed information
is provided. However the solution of the R-H problem as a rule
involves the presence of apparent singularities which in the case at
hand can be chosen to be no more than one. Relaxing the R-H data
always keeping the trace of the five original monodromies fixed
increases the number of degrees of freedom and counting the
degrees of freedom makes one to expects that for any set of $\delta$'s
the problem is soluble but we have no rigorous proof of that. If a
positive answer is given to such a question the problem naturally
arises of the uniqueness of the solution.  Uniqueness combined with
the rigorous results of section \ref{green} would imply analyticity in
the full circle of radius $1$. At $x=1$ we have unavoidably a
singularity as the monodromy contour is pinched by two singularities.

\section*{Appendix 1}

In this appendix we report in orderly way to the reader's benefit
the first three coefficients of the expansion of $C(x)$ in $x$
\begin{equation}
  C(0)=\delta_\nu-\delta_0-\delta
\end{equation}
\begin{equation}
C'(0) = 2\delta -\frac{(2\delta
  +C(0))(\delta_\infty-\delta_1+\delta_\nu)}{2\delta_\nu}
\end{equation}
\begin{eqnarray}
& &C''(0)=-\frac{(\delta_\infty+\delta_\nu-\delta_1)
[C'(0)-3\delta+b_{0,1}^2(2\delta_\nu+\delta_\infty-\delta_1)]}{\delta_\nu}\\
&-&\frac{(C(0)+ 3\delta-3b_{0,1}^2\delta_\nu)
[3\delta_1^2+3\delta_\nu^2+3\delta_\infty(1+\delta_\infty)+
\delta_\nu(3+2\delta_\infty)-3\delta_1(1+2\delta_\nu+2\delta_\infty)]}
{\delta_\nu(3+4\delta_\nu)}\nonumber
\end{eqnarray}
where 
\begin{equation}
  b_{0,1} =\frac{2\delta+C(0)}{2\delta_\nu}
\end{equation}
\begin{eqnarray}
& &C'''(0)=
((-C(0) - 4\delta + 10 b_{1,1} b_{0,1} \delta_\nu + 4 b_{0,1}^3\delta_\nu)
   (-6 \delta_1 + 6 \delta_\nu - 6\delta_\infty))/(6 (2 +
\delta_\nu))\nonumber\\
 &+& 
 (-6 \delta_1 + 6 \delta_\nu - 6\delta_\infty)
  (-((-C(0) - 4 \delta + 10 b_{1,1} b_{0,1} \delta_\nu + 4 b_{0,1}^3
\delta_\nu)\nonumber\\
& &   (12 - 5 \delta_1 + \delta_\nu + 5 \delta_\infty))/(6 (-3 - 4\delta_\nu)
     (2 + \delta_\nu)) \nonumber\\
&+& (C'(0) - 4 \delta - 4 b_{1,1} b_{0,1} \delta_1 - 
     b_{0,1}^3 \delta_1 + 
     6 b_{1,1} b_{0,1} \delta_\nu + 3 b_{0,1}^3 \delta_\nu - 6 b_{0,1} b_{0,2}
     \delta_\nu + 4 b_{1,1} b_{0,1} \delta_\infty + 
     b_{0,1}^3 \delta_\infty)/\nonumber\\
&&     (-3 - 4 \delta_\nu))\nonumber\\ 
&+& (-6 \delta_1 + 6 \delta_\nu + 6 \delta_\infty)
  (((-C(0) - 4\delta + 10 b_{1,1} b_{0,1} \delta_\nu + 4 b_{0,1}^3\delta_\nu)
     (-6 \delta_1 + 2 \delta_\nu - 2 \delta_\infty))/\nonumber \\
     &&(24 \delta_\nu 
    (2 + \delta_\nu))\nonumber\\
 &-& (C''(0) + 2 b_{1,1} b_{0,1} \delta_1 - 4 b_{0,1} b_{0,2} \delta_1 
+ 2 b_{1,1} b_{0,1} \delta_\nu + 
     8 b_{0,1} b_{0,2} \delta_\nu - 2 b_{1,1} b_{0,1} \delta_\infty 
  + 4b_{0,1} b_{0,2}\delta_\infty)/(4\delta_\nu)\nonumber\\ 
&+& ((6 - 6\delta_1 + 2 \delta_\nu + 6\delta_\infty)
  (-((-C(0) - 4\delta + 10 b_{1,1} b_{0,1}\delta_\nu + 4b_{0,1}^3
  \delta_\nu)\nonumber\\
& & (12 - 5\delta_1 + \delta_\nu + 5\delta_\infty))/(6(-3 - 4\delta_\nu)
        (2 + \delta_\nu))\nonumber\\ 
&+& (C'(0) - 4\delta - 4 b_{1,1} b_{0,1} \delta_1 - 
      b_{0,1}^3\delta_1 + 
      6 b_{1,1} b_{0,1}\delta_\nu + 3 b_{0,1}^3\delta_\nu - 6 b_{0,1}
      b_{0,2}\delta_\nu + 
        4 b_{1,1}b_{0,1}\delta_\infty + b_{0,1}^3\delta_\infty)/\nonumber\\
& &(-3 - 4\delta_\nu)))/(4\delta_\nu))
\end{eqnarray}
where
\begin{equation}
b_{1,1} =\frac{3 \delta - 3 b_{0,1}^2 \delta_\nu+C(0)}{3+4 \delta_\nu} 
\end{equation}
and 
\begin{eqnarray}
&&  b_{0,2} =\bigg[(\delta_1+\delta_\nu-\delta_\infty)
    \big((5\delta_\nu-3)b_{0,1}^2- 9 \delta\big)+ \nonumber\\
&&    
        (3-3\delta_1+3\delta_\infty+\delta_\nu) C(0)+
    (3+4\delta_\nu)C'(0)\bigg]\bigg\slash
\big(2\delta_\nu(3+4\delta_\nu)\big)~.
\end{eqnarray}

\section*{Appendix 2}

In the perturbative computation of eq.(\ref{firstorder}) and higher
order computation integrals of the type
\begin{equation}  
\int_1^\infty t_j\frac{1}{z^m(z-1)} t_k dz
\end{equation}
intervene.
All such integrals can be computed
exactly in terms of hypergeometric functions and derivative thereof.

The integral of $1/(z^2(z-1))$ can be computed by taking the variation
w.r.t. $\delta_\nu$.

\begin{equation}
\dot t''+\dot Q_0 t+ Q_0 \dot t=0,~~~~\dot Q_0 = -\frac{1}{z^2(z-1)}
\end{equation}
with
\begin{equation}
\dot t = -\frac{2}{\lambda_\nu}\frac{\partial t}{\partial \lambda_\nu}~.
\end{equation}
Then
\begin{equation}
  \int_1^z t_k \dot Q_0 t_j dz =
  t_k' \dot t_j - t_k \dot t'_j\bigg|_1^z~.
\end{equation}
The contribution at $z=1$ vanishes and as we have asymptotically
\begin{equation}
  t_k\approx B_{k1} z^\frac{1-\lambda_\infty}{2}+
  B_{k2} z^\frac{1+\lambda_\infty}{2}
\end{equation}
with $B_{kl}$ given by the matrix (\ref{Bmatrix}) the result is
\begin{equation}\label{intzm2}
  \int_1^\infty t_k \dot Q_0 t_j dv = -\int_1^\infty t_k
  \frac{1}{z^2(z-1)} t_j dv = -\lambda_\infty B_{kl} \dot
  B_{Jim}\varepsilon_{lm}
\end{equation}
with $\varepsilon_{lm}$ the antisymmetric symbol.
All the others can be computed recursively from the variation
\begin{equation}
z= \frac{v-\varepsilon/v^{N-1}}{1-\varepsilon}
\end{equation}
which leaves $z=1$ and $z=\infty$ fixed. The result is
with
\begin{equation}
\tilde Q_0(v) = Q_0(z(v)) (z'(v))^2-\{z,v\}
\end{equation}
\begin{equation}
  \dot{\tilde Q}(v)= \frac{1}{v-1}\bigg(\sum_{l=1}^NA^{(N)}_{l,N}\frac{1}{v^{2+l}}-
  (2\delta_1+\delta_2)\frac{1}{v^2}\bigg)
\end{equation}
where $A^{(N)}$ is the matrix given in (\ref{Acolumn})
and
\begin{equation}
  \dot{\tilde t}_k = -\frac{1}{2}\big(1+\frac{N-1}{v^N})t_k(v)+
  \big(v-\frac{1}{v^{N-1}}\big) t_k'(v)
\end{equation}
and thus
\begin{equation}\label{derfunction}
  \int_1^v t_k \dot{ \tilde Q} t_l dv =
  t_k' \dot t_j - t_k \dot t'_j\bigg|_1^v~.
\end{equation}
The integral
\begin{equation}
\int_1^z t_k\frac{1}{z^2(z-1)}t_j dz
\end{equation}
has been given in (\ref{intzm2}). The integral
\begin{equation}
\int_1^z t_k\frac{1}{z^3(z-1)}t_j dz
\end{equation}
is computed by using the previous and the result obtained with $N=1$
and iteratively all the others.

In computing higher orders of perturbation theory one needs also matrix
elements of the variation of the $y_j$. These can obtained by
pushing the above procedure to higher order. E.g. to order
$\varepsilon^2$ we have
\begin{equation}
\ddot t''+\ddot Q t +2 \dot Q \dot t + Q \ddot t =0
\end{equation}
from which
\begin{equation}0={\ddot t_j}'t_k- {\ddot t_j}t'_k\big|^z_1+
\int_1^z (t_j\ddot Q t_k+2\dot t_j \dot Q t_k)dz  
\end{equation}
thus providing 
\begin{equation}
\int_1^z\dot t_j \dot Q t_k dz  
\end{equation}
\bigskip
which gives through (\ref{derfunction}) the matrix element with the
first derivative.

We remark that the algebraic procedure of section
\ref{algebraic} provides the expansion of $C(x)$ without the need of
computing these perturbative integrals.

\vfill


\end{document}